\documentclass[conference,10pt,twocolumn]{IEEEtran}
\usepackage{amssymb}
\usepackage{algorithm}
\usepackage{algpseudocode}
\usepackage{algorithmicx}
\usepackage{amsfonts}
\usepackage{amsmath}
\usepackage{amssymb}
\usepackage{amsthm}
\usepackage{array}
\usepackage{bbding}
\usepackage{bigints}
\usepackage{booktabs}
\usepackage{cite}
\usepackage{cleveref}
\usepackage{color}
\usepackage{diagbox}
\usepackage{epsfig,latexsym}
\usepackage{epstopdf}
\usepackage{graphicx}
\usepackage{fancyhdr}
\usepackage{float}
\usepackage{flushend}
\usepackage{indentfirst}
\usepackage{lastpage}
\usepackage{makecell}
\usepackage{mathtools}
\usepackage{multirow}
\usepackage{pifont}
\usepackage{psfrag}
\usepackage{setspace}
\usepackage{stfloats}
\usepackage{subfloat}
\usepackage{times}
\usepackage{xcolor}
\usepackage{subfigure}

\begin{document}
\title{\LARGE OTFS-based Integrated Positioning and Communication Systems with Low-Resolution ADCs}

\author{{Yueyi Yang}, {Zeping Sui}, {Zilong Liu}, {and Leila Musavian}\\
School of Computer Science and Electronics Engineering, University of Essex, Colchester CO4 3SQ, U.K.\\
Emails: \{yy24071,zilong.liu,leila.musavian\}@essex.ac.uk, zepingsui@outlook.com
\thanks{}
%
}
\maketitle

\begin{abstract}

This paper proposes a two-phase orthogonal time–frequency space (OTFS)-based integrated positioning and communication (IPAC) framework under realistic low-resolution analog-to-digital converters (ADCs). In the uplink phase, the positioning signal is used to estimate channel parameters, which are subsequently used to determine the user's position. The spatial smoothing-multiple signal classification algorithm is introduced to estimate the angle-of-arrival, whereas an iterative interference cancellation scheme is conceived for the remaining parameters' estimation. The corresponding Cramer–Rao lower bounds of channel parameters and user position are also derived. During the downlink communication phase, the estimated parameters are exploited to improve beamforming at the base station. Simulation results evaluate the impact of ADC quantizer resolutions. Specifically, it is shown that enhanced downlink bit error rate performance can be achieved with improved uplink positioning, while the use of low-resolution ADCs induces noticeable performance degradation in the OTFS-IPAC system.


\end{abstract}
\IEEEpeerreviewmaketitle

\section{Introduction}\label{Section1}
Next-generation wireless networks are expected to support both high-rate communication and accurate positioning in emerging scenarios \cite{9779322,sui2025multi}, including autonomous driving, unmanned aerial platforms, indoor navigation, and industrial automation, thereby driving the evolution towards integrated positioning and communication (IPAC). In contrast to conventional designs, where communication and positioning are implemented separately, IPAC integrates both functionalities within a unified system for improved spectral efficiency and lower hardware costs \cite{11184827}.

From a waveform perspective, conventional orthogonal frequency-division multiplexing (OFDM) is widely adopted in modern communication systems. However, its error rate performance generally suffers from severe inter-carrier interference imposed by large Doppler, carrier frequency offsets, and timing synchronization errors in high-mobility scenarios \cite{wei2021orthogonal,10061469}. In recent years, orthogonal time–frequency space (OTFS) modulation has emerged as a promising alternative for sending data symbols in the delay-Doppler (DD) domain \cite{11185315,10250854,10129061}. Compared to OFDM, OTFS is Doppler resilient, thanks to its quasi-static and sparse DD-domain channel matrix, which is a consequence of its unique signal structure \cite{10183832}. From this standpoint, OTFS is a suitable waveform candidate for IPAC system design operating in high-mobility scenarios. Nevertheless, literatures on OTFS-based IPAC is limited to \cite{10274474,11079715}. Explicitly, in \cite{10274474}, a low-complexity 3D-FFT-based algorithm was proposed to estimate the delay, Doppler, and angle-of-arrival (AoA) for massive MIMO-OTFS systems. This approach utilizes multidimensional Fourier processing to achieve efficient parameter estimation, making it suitable for real-time implementations. In addition, the Cramer–Rao lower bounds (CRLBs) were derived to describe the limits of delay and Doppler estimation under finite time–frequency resources, providing useful benchmarks for algorithm design \cite{9524512}. Considering the reconfigurable intelligent surface-assisted OTFS systems with limited Global Positioning System availability, the authors in \cite{11079715} investigated the joint velocity and position estimation algorithm, which combines multiple signal classification (MUSIC)-based AoA estimation with Kalman filtering. Moreover, a low-Peak to Average Power Ratio frame structure was designed to embed velocity information in the presence of noise and interference.

That said, the above papers generally assume ideal hardware, thus overlooking the practical constraints of wideband and large-scale MIMO systems. Moreover, from the system energy efficiency point of view, the power consumption of analog-to-digital converters (ADCs) increases exponentially with the quantizer resolutions~\cite{7886292,8320852,9739838,8625478}. Against this problem, low-resolution ADCs are advocated as a practical and power-efficient approach. Specifically, the additive quantization noise model (AQNM) \cite{7886292} was adopted to represent the output of a low-resolution ADC as a scaled input signal with an additive distortion term, whose covariance is determined by the quantization resolution. A Bayesian data detection framework based on a generalized expectation-consistent signal recovery technique was developed in \cite{8320852}, yielding a minimum mean-square-error (MMSE) symbol estimator. By exploiting the structure of the MIMO-OFDM channel matrix, a low-complexity algorithm relying on FFT operations and matrix–vector multiplications was derived. Moreover, a quantized turbo OFDM receiver involving joint channel estimation and data detection was designed in \cite{9739838} to mitigate the nonlinear ADC distortion. The authors in \cite{8625478} investigated the delay-domain sparse channel estimation and data detection for massive MIMO-OFDM systems with low-resolution ADCs. Corresponding Bayesian CRLBs were derived, and a variational Bayes–based sparse recovery algorithm with soft symbol decoding was developed to enable iterative data-aided channel estimation.

In the literature, there are only a few papers that consider low-resolution ADCs for OTFS systems. In \cite{10758799}, a variational Bayesian learning-based sparse channel estimation scheme for the MIMO-OTFS system was proposed by leveraging channel sparsity in the DD-domain. It is worth noting that the aforementioned studies primarily focused on pure communication systems. Against this backdrop, we propose an OTFS-IPAC framework with realistic low-resolution ADCs. Our contributions are summarized as follows:
\begin{itemize}
	\item We first develop an OTFS-based IPAC framework in the delay–Doppler–angular (DDA) domain using low-resolution ADCs. In the uplink phase, based on random positioning OTFS signals, we present a spatial smoothing-MUSIC (SS-MUSIC) algorithm \cite{thompson2002performance} along with a path-wise interference cancellation algorithm to estimate parameters. In the downlink stage, the above estimated parameters enable a pre-equalization strategy based on zero-forcing (ZF) beamforming at the base station (BS), through which uplink parameter estimation facilitates downlink communication performance.
	\item We derive the CRLBs of channel parameters. By exploiting the geometric relationship between channel parameters, we further obtain the CRLB of user position, thereby revealing the fundamental estimation limits of the proposed OTFS-IPAC system under low-resolution ADCs.
	\item Our extensive simulation results demonstrate that 1) Lower downlink communication BER can be achieved by improving the uplink positioning performance; 2) Exploiting low-resolution ADCs leads to the performance degradation in parameter estimation, user positioning, and communication.
\end{itemize}
\section{System Model}\label{Section2}
Let us consider an OTFS-IPAC system comprising one BS and a single-antenna user terminal, utilizing the time division duplexing (TDD) scheme, as illustrated in Fig. 1(a). Specifically, the BS is equipped with $N_t$ transmit antennas (TAs) and $N_r$ receive antennas (RAs). Moreover, the user terminal transmits uplink positioning signals to the BS during the $t$-th timeslot, while the BS transmits downlink communication signals to the user terminal in the $(t+1)$-th timeslot. 
\begin{figure}[htbp]
	\centering
	\includegraphics[width=0.6\linewidth]{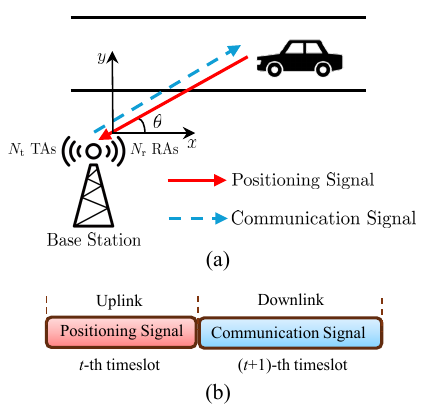}
	\caption{Illustration of (a) the proposed OTFS-IPAC systems with uplink positioning and downlink communication signals, (b) TDD-based uplink positioning and downlink communication signal frame structure. }
	\label{Figure1}
\end{figure}

\subsection{OTFS Modulation}
Consider a DD-domain symbol matrix $\mathbf{X}\in \mathbb{C}^{M \times N}$, where $M$ is the number of subcarriers and $N$ is the number of time-slots. Then, the time–frequency (TF)-domain symbol matrix can be obtained via the inverse symplectic finite Fourier transform (ISFFT), yielding $\mathbf{X}_{\mathrm{TF}} = \mathbf{F}_M \mathbf{X} \mathbf{F}_N^{H}$, where $\mathbf{F}_M$ denotes the normalized discrete Fourier transform (DFT) matrices of size $M$, and $(\cdot)^{H}$ denotes the Hermitian transpose operator. Upon using the Heisenberg transform with a rectangular transmit pulse and appending a single cyclic prefix (CP) to the entire OTFS frame, the transmit symbol matrix in the time-domain (TD) is given by
\begin{align}
	\mathbf{S} = \mathbf{I}_{M} \mathbf{F}_M^H \mathbf{X}_{\text{TF}} = \mathbf{I}_{M} \mathbf{X} \mathbf{F}_N^H.
\end{align}
Then, the transmit TD symbol vector can be obtained as $\mathbf{s} = \operatorname{vec}(\mathbf{S}) = \left( \mathbf{F}_N^H \otimes \mathbf{I}_{M} \right) \mathbf{x}$, where $\mathbf{x} = \text{vec}(\mathbf{X}) \in \mathbb{C}^{MN \times 1}$, $\operatorname{vec}(\cdot)$ denotes the vectorization operator, and $\otimes$ denotes the kronecker product oprator.
The DDA-domain channel impulse response can be modelled as
\begin{align}\label{}
	\mathbf{h}(\tau, \nu, \theta) = \sum_{p=0}^{P-1} h_p \mathbf{a}_r^H(\theta)\delta(\tau - \tau_p) \delta(\nu - \nu_p) \delta(\theta - \theta_p),
\end{align}
where $P$ is the number of paths, and $\delta(\cdot)$ represents the Dirac-delta function. Each path is characterized by complex-valued gain $h_p$, delay $\tau_p$, Doppler $\nu_p$, and AoA $\theta_p$. Here we consider a uniform linear array (ULA) at the BS with $N_r$ elements with inter-element spacing as half of the wavelength $d = \lambda/2$, and the receive steering vector is given by $\mathbf{a}_r(\theta) = \left[ 1, e^{\jmath\pi \sin\theta}, \ldots, e^{\jmath\pi (N_r - 1) \sin\theta} \right]^{T}$, where $(\cdot)^T$ denotes the transpose operator. Moreover, the delay and Doppler shifts of the $p$th path can be modeled as
\begin{align}\label{}
	\tau_p = \frac{l_{p}}{M \Delta f}, \quad \nu_p = \frac{k_{p}+\kappa_p}{N T},
\end{align}
where $T$ denotes the symbol duration, $\Delta f$ refers to the subcarrier spacing, and $l_p$ and $ k_p$ are normalized integer delay and Doppler indices, respectively, while $\kappa_p\in[-1/2,1/2]$ denotes the normalized fractional Doppler shifts.  We assume that the normalized delays are given by $\mathcal{L}=\{l_0,l_1,\ldots,l_{P-1}\}=\{1,2,\ldots,P\}$ \cite{9524512}.

\subsection{Received Signals with Low-resolution ADCs}
The received signal in TD can be expressed as
\begin{align}\label{}
	\mathbf{r}(t) = \sum_{p=0}^{P-1} h_p e^{\jmath2\pi \nu_p (t - \tau_p)}  \mathbf{a}_r(\theta_p) s(t - \tau_p) + \tilde{\mathbf{w}}(t),
\end{align}
where $\tilde{\mathbf{w}}(t)$ denotes the complex additive white Gaussian noise with mean of zero and variance of $\sigma^2$. After sampling at $\{t=nT_s,n=0,\ldots,N-1\}$ with $T_s=1/(M\Delta f)$ and discarding the CP, the $N_r$-dimensional received signal can be formulated in a vector form as
\begin{align}\label{r_matrix}
	\mathbf{r} = \mathbf{H} \mathbf{s} + \tilde{\mathbf{w}},
\end{align}
where the effective channel matrix $\mathbf{H}\in \mathbb{C}^{MNN_r \times MNN_t}$ can be formulated as
\begin{align}\label{TD channel}
	\mathbf{H}  = \sum_{p=0}^{P-1} h_p 
	\left[ 
	\bigl( \mathbf{a}_r(\theta_p)  
	\otimes 
	\bigl( \boldsymbol{\Pi}^{l_p} \boldsymbol{\Delta}^{k_p+\kappa_p} \bigr)
	\right],
\end{align}
where $\boldsymbol{\Pi}$ denotes the forward cyclic shift matrix, and the Doppler shifts are captured by $\boldsymbol{\Delta} \triangleq 
\mathrm{diag}\!\big\{ 1, e^{\jmath \frac{2\pi}{MN}}, \dots, e^{\jmath\frac{2\pi(MN-1)}{MN}} \big\}$.

In this work, we utilize AQNM to model the effect of low-resolution ADCs, which is represented by a scaling gain with an additive quantization noise term. Consequently, the TD receive signal with low-resolution ADCs can be formulated based on \eqref{r_matrix} as
\begin{align}\label{r_ad}
	\mathbf{r}_{\mathrm{ad}} 
	= \mathbb{Q} (\mathbf{r}) = \alpha \mathbf{H}\mathbf{s} + \alpha\tilde{\mathbf{w}} + \mathbf{w}_{\mathrm{ad}},
\end{align}
where $\alpha=1 - \beta$ is the scaling quantization gain, which is dependent on the quantization bits $b$. The values of $\beta$ for low-resolution ADCs are shown in Table~\ref{beta_values}. Moreover, the quantization noise $\mathbf{w}_{\mathrm{ad}}$ follows the complex Gaussian distribution with zero mean and covariance matrix $\mathbf{C}_{\mathrm{ad}}= \alpha\beta\,\mathrm{diag}\!\left(\mathbb{E}[\mathbf{r}\mathbf{r}^H]\right)$, i.e., we have $\mathbf{w}_{\mathrm{ad}} \sim \mathcal{CN}(0, \mathbf{C}_{\mathrm{ad}})$. As expressed in \eqref{r_ad}, low-resolution ADC degrades the received signal power through the scaling factor $\alpha$, while the effective noise $\bar{\mathbf{w}}_{\mathrm{ad}} = \alpha \tilde{\mathbf{w}} + \mathbf{w}_{\mathrm{ad}}$ includes not only channel noise but also distortion originating from the limited ADC resolutions.

Consequently, the TD received sample matrix is given by $\mathbf{R}_{\mathrm{ad}} =  \operatorname{vec}^{-1}(\mathbf{r}_{\mathrm{ad}})\in\mathbb{C}^{M\times N_rN}$. Based on the Wigner transform and SFFT operation, the DD-domain signal can be obtained as
 \begin{align}\label{}
 	\mathbf{Y}_{\mathrm{ad}}  = \mathbf{F}_M^H(\mathbf{F}_M \mathbf{R}_{\mathrm{ad}})(\mathbf{I}_{N_r} \otimes \mathbf{F}_N)
 	            = \mathbf{I}_M\mathbf{R}_{\mathrm{ad}} (\mathbf{I}_{N_r} \otimes \mathbf{F}_N).
 \end{align}
The vectorized form of the received signal in the DDA-domain $\mathbf{y}_{\mathrm{ad}}= \operatorname{vec}(\mathbf{Y}_{\mathrm{ad}})= \left( \mathbf{I}_{N_r} \otimes \mathbf{F}_N \otimes \mathbf{I}_{M} \right)\mathbf{r}_{\mathrm{ad}}$ can be obtained as
\begin{equation}
\begin{aligned}
	\mathbf{y}_{\mathrm{ad}} &=\alpha \mathbf{G}_\text{U}\mathbf{x} + \mathbf{w},
\end{aligned}
\end{equation}
where $\mathbf{G}_\text{U} = \left( \mathbf{I}_{N_r} \otimes \mathbf{F}_N \otimes \mathbf{I}_{M} \right) 
	\mathbf{H} \left( \mathbf{F}_N^{H} \otimes \mathbf{I}_M \right)$ denotes the uplink effective channel matrix in the DDA-domain, and the corresponding noise vector is given by $\mathbf{w} =\alpha \left( \mathbf{F}_N \otimes\mathbf{I}_M \right) \tilde{\mathbf{w}} 
        + \left( \mathbf{F}_N \otimes\mathbf{I}_M \right) \mathbf{w}_{\mathrm{ad}}$.

 \begin{table}[t]
\footnotesize
	\centering
	\caption{Values of $\beta$ for small $b\leq 5$ \cite{ma2023scalable}}
	\label{beta_values}
	\begin{tabular}{c|ccccc}
		\hline
		$b$ & 1 & 2 & 3 & 4 & 5 \\
		\hline
		$\beta$ & 0.3634 & 0.1175 & 0.03454 & 0.009497 & 0.002499 \\
		\hline
	\end{tabular}
    \vspace{-2em}
\end{table}
\section{CRLB Derivation with Low-resolution ADCs} \label{Section3}
In this section, we derive the CRLB to evaluate the performance of channel parameters and position estimation in our proposed OTFS-IPAC systems. According to \eqref{r_ad}, the covariance matrix $\bar{\mathbf{w}}_{\mathrm{ad}} = \alpha \mathbf{w} + \mathbf{w}_{\mathrm{ad}}$ can be formulated as 
\begin{align}\label{}
\mathbf{\Sigma} = \alpha^2\mathbf{C}_w+\mathbf{C}_{\mathrm{ad}},
\end{align}
where $\mathbf{C}_w$ denotes the covariance matrix of the addictive white Gaussian noise vector $\mathbf{w}$, while $\mathbf{C}_{\mathrm{ad}}$ characterizes the covariance matrix of the quantization noise generated by the low-resolution ADCs. Let $\mathcal{I} = \{h_0,\ldots,h_{P-1},\theta_0,\ldots,\theta_{P-1},k_{0}+\kappa_0,\ldots,k_{P-1}+\kappa_{P-1}\}$ denotes the unknown channel parameter set with a size of $3P$. Based on the input-output relationship of \eqref{r_ad}, the logarithm probability density function (PDF) of $\mathbf{r}_{\mathrm{ad}}$ conditioned on $\mathcal{I}$ can be written as
\begin{equation}
\begin{aligned}\label{17}	
	f &\triangleq \ln p(\mathbf{r}_{\mathrm{ad}}, \mathcal{I})	\\
	&= -(\mathbf{r}_{\mathrm{ad}} -  \alpha \mathbf{H}\mathbf{s})^H
	\mathbf{\Sigma}^{-1}
	(	\mathbf{r}_{\mathrm{ad}} -  \alpha \mathbf{H}\mathbf{s}) - \frac{1}{2} \ln \det(\mathbf{\Sigma}).
	\mathcal{}
\end{aligned}
\end{equation}
To obtain the CRLB of channel parameters, we first calculate the Fisher information matrix (FIM) by taking the second-order derivatives of the log-likelihood function $f$. Specifically, the FIM can be formulated as a function of the channel parameter set $\mathcal{I}$, yielding
\begin{align}\label{12}
	\mathbf{J}(\mathcal{I}) = [\mathrm{J}_{i,j}]_{1 \le i,j \le 3P}, \qquad
	\mathrm{J}_{i,j} = -\mathbb{E}\!\left( \frac{\partial^2 f}{\partial \mathcal{I}_i \partial \mathcal{I}_j^*} \right),
\end{align}
where $\mathrm{J}_{i,j}$ denotes the $(i,j)$th elements of the matrix $\mathbf{J}(\mathcal{I})$. By substituting the Gaussian likelihood of \eqref{17} into \eqref{12}, the FIM elements can be expressed as
\begin{align}\label{}
	\mathrm{J}_{i,j} 
	&=
	\begin{cases}
		\left(\alpha\frac{\partial \mathbf{H}}{\partial {\mathcal{I}}_i} \mathbf{s}\right)^H
		\mathbf{\Sigma}^{-1}
		\left(\alpha\frac{\partial \mathbf{H}}{\partial {\mathcal{I}}_j} \mathbf{s}\right), &i,j\in[1,P],
		\\[2ex]
		-2 \mathrm{Re} \left\{
		-\left(\alpha
		\frac{\partial \mathbf{H}}{\partial {\mathcal{I}}_j} {\mathbf{s}}
		\right)^H
		\mathbf{\Sigma}^{-1}
		\left(\alpha\frac{\partial \mathbf{H}}{\partial {\mathcal{I}}_i} {\mathbf{s}}\right)
		\right\},&\text{otherwise}.
	\end{cases}	
\end{align}
Then, we compute the derivatives of $\mathbf{H}$ with respect to each channel parameter. For the $p$-th path, the partial derivative 
$\frac{\partial \mathbf{H}}{\partial {\mathcal{I}}_p}$ can be calculated as
\begin{align}\label{}
	\frac{\partial \mathbf{H}}{\partial h_p}&=\mathbf{a}_r(\theta_p) \otimes \boldsymbol{\Pi}^{l_p} \boldsymbol{\Delta}^{k_p+\kappa_p},\nonumber\\
	\frac{\partial \mathbf{H}}{\partial \theta_p}&=h_p [\mathbf{d}_r \odot \mathbf{a}_r(\theta_p) ] \otimes \boldsymbol{\Pi}^{l_p} \boldsymbol{\Delta}^{k_p+\kappa_p},\nonumber\\
	\frac{\partial \mathbf{H}}{\partial k_p}&=h_p \mathbf{a}_r(\theta_p)\otimes \boldsymbol{\Pi}^{l_p}\mathbf{D}
	\boldsymbol{\Delta}^{k_p+\kappa_p},
    \end{align}
with
\begin{align}
   \mathbf{d}_r &\triangleq	\jmath\pi\cos(\theta_p)[0,1,...,N_{r}-1]^T,\nonumber\\
	\mathbf{D} &\triangleq \mathrm{diag}\left\{\jmath\frac{2\pi}{MN} [0,1,...,MN-1]^T\right\}, 
\end{align}
where $\odot$ denotes the Hadamard product operator. By inverting $\mathbf{J}(\mathcal{I})$, we can obtain the CRLBs for the channel parameters. The diagonal elements of $\mathbf{J}(\mathcal{I})$ provide the minimum achievable variances of any unbiased estimator for the corresponding channel parameters.

Considering that only the line-of-sight (LoS) path carries reliable positioning information, the user position is inferred directly from the estimated LoS channel parameters. For convenience, we assume the BS is located at the origin, and the user position vector $\mathbf{u} = [x,y]^T$ can be expressed as
\begin{align}\label{}
		x = c \tau_0 \mathrm{cos} {\mathrm{\theta}}_0, \ y = c\tau_0 \mathrm{sin} {\mathrm{\theta}}_0,
\end{align}
where $\tau_0$ and $\theta_0$ denote the delay and AoA of the LoS path, respectively, and $c$ is the speed of light.
The FIM of $\mathbf{u}$ can be formulated as
\begin{equation}
	\mathrm{J}(\mathbf{u})
	= \nabla_{\mathbf{u}}^{{T}} \theta_0 \, \mathrm{J} (\theta_0) \, \nabla_{\mathbf{u}} \theta_0,
\end{equation}
where $\mathrm{J} (\theta_0)$ denotes the FIM of the LoS path's AoA, and the Jacobian matrix $\nabla_{\mathbf{u}}\theta_0$ can be derived as
\begin{align}\label{J_p}
 	\nabla_{\mathbf{u}} \theta_0 = 
 		\frac{\mathbf{z}^{T}}{\sqrt{1-(\boldsymbol{\eta}_{\mathbf{u}}^{T}\mathbf{z})^2}} \big( \mathbf{I} - \boldsymbol{\eta}_{\mathbf{u}}\boldsymbol{\eta}_{\mathbf{u}}^{T} \big) / d,
\end{align}
where $d = \|\mathbf{u} - \mathbf{u}_\text{BS}\| = \|\mathbf{u} \|$ refers to the distance between the BS and the user terminal, and $\boldsymbol{\eta}_{\mathbf{u}} = \mathbf{u}/d$ is the direction vector. Moreover, $\mathbf{z} = [0,1]^T$ denotes a unit vector which is not aligned with the user's velocity or $\boldsymbol{\eta}_{\mathbf{u}}$. Finally, the CRLB of the user position is derived as 
\begin{align}\label{}
	\text{CRLB}(\mathbf{u}) =  \mathrm{J}^{-1}(\mathbf{u}),
\end{align}
which represents the minimum achievable MSE of any unbiased position estimator in the proposed OTFS-IPAC system.
\section{Signal Processing of OTFS-IPAC Systems}\label{Section4}
\subsection{Uplink Positioning}
It can be observed from \eqref{TD channel} that the angular information is embedded in the array steering vector $\mathbf{a}_r(\theta_p)$ and is decoupled from delay and Doppler shifts. Therefore, the AoAs can be estimated only from the spatial covariance using the MUSIC algorithm, which is independent of delay and Doppler parameters. However, the multi-path propagation effect introduces correlation among receive signals, which degrades the performance of conventional subspace-based estimators. To mitigate this effect, we employ the SS-MUSIC algorithm \cite{thompson2002performance}, which suppresses the inter-path correlation via spatial smoothing. By reconstructing an equivalent uncorrelated covariance matrix, SS-MUSIC enables high-resolution AoA estimation under strong multi-path coupling. As shown in Fig. 2, for a ULA with $N_r$ RAs, we partition the array into $K$ overlapping subarrays of length $L$, i.e., we have $K = N_r - L +1$. 
Then, the smoothed covariance matrix can be formulated as
\begin{align}\label{eq_Rss}
	\mathbf{R}_{ss}  = \frac{1}{K} \sum_{k=1}^{K}  \mathbf{R}_{k} = \frac{1}{KL} \sum_{k=1}^{K} \sum_{l=1}^{L} \mathbf{Y}_{l}\mathbf{Y}_{l}^H,
\end{align}
where $\mathbf{Y}_l$ represents the received signal vector corresponding to the $l$-th antenna element within the $k$-th  subarray. By applying singular value decomposition to the covariance matrix, \eqref{eq_Rss} can be rewritten as
\begin{align}
	\mathbf{R}_{ss}  = \mathbf{E}_s \mathbf{\Lambda}_s \mathbf{E}_s^{H} + \mathbf{E}_n \mathbf{\Lambda}_n \mathbf{E}_n^{H},
\end{align}
where $\mathbf{E}_s$ and $\mathbf{E}_n$ denote the signal and noise subspaces, respectively, and $\boldsymbol{\Lambda}_s$ and $\boldsymbol{\Lambda}_n$ are diagonal matrices containing the corresponding signal and noise eigenvalues. The MUSIC pseudo-spectrum is obtained as
\begin{align}\label{}
	Q(\theta) = \frac{1}{\mathbf{a}_r^{H}(\theta) \, \mathbf{E}_n \, \mathbf{E}_n^{H} \, \mathbf{a}_r(\theta)},
\end{align}
where the spectral peaks appear at the AoAs of the echoes.

Next, we present an iterative algorithm to estimate other parameters. For the $p$-th path, the Doppler shift is estimated by maximizing the correlation metric
\begin{align}\label{DD phase1}
	\hat k_p = \arg\max_{k_p\in\Omega} 
	\big|\alpha\,(\Gamma_p\mathbf{x})^H \mathbf{y}_r\big|^2,
\end{align}
\begin{figure}[tbp]
	\centering
	\includegraphics[width=0.7\linewidth]{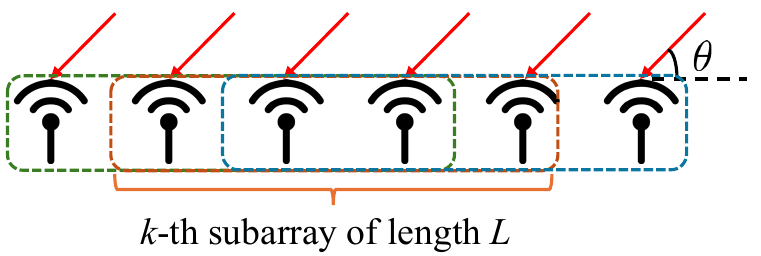}
	\caption{A toy example of SS-MUSIC using the receive array with $N_r=6, L=4$, and $K=6$. }
	\label{Figure2}
	\vspace{-1em}
\end{figure}
where $\mathbf{y}_r = \mathbf{y} - \alpha\sum_{q=1}^{p-1} \hat{h}_q \mathbf{\Gamma}_q(\hat{\theta}_q, \hat{k}_q) \mathbf{x}$ denotes the residual signal after interference cancellation of the previously detected components. The search set is defined as 
$\Omega = \big\{\tfrac{k}{NT} \,\big|\, k=-\tfrac{N}{2},\ldots,\tfrac{N}{2}-1 \big\}$. Moreover, the atom dictionary $\mathbf{\Gamma}_p(\theta_p,k_p)$ incorporates the array response and the Doppler shifts, yielding
\begin{align} \label{Gamma}
	\Gamma_p(\theta_p,k_p) 
	= \mathbf{a}_r(\theta_p)\otimes (\mathbf{F}_N\otimes \mathbf{I}_M)\,
	\mathbf{\Pi}^{l_p}\mathbf{\Delta}^{k_p}(\mathbf{F}_N^H\otimes \mathbf{I}_M).
\end{align}
Then, the initial coarse search over $\Omega$ is first carried out, followed by a refinement of the fractional Doppler through a golden section search \cite{10061469}. Once $\hat k_p$ is obtained, the corresponding channel gain is computed in closed form as
\begin{align}\label{channel gain}
	\hat h_p = 
	\frac{\big[\Gamma_q(\hat\theta_q,\hat k_q+\hat\kappa_q)\mathbf{x}\big]^H \mathbf{y}_r}
	{\mathbf{x}^H \Gamma_q^H(\hat\theta_q,\hat k_q+\hat\kappa_q)\Gamma_q(\hat\theta_q,\hat k_q+\hat\kappa_q)\mathbf{x}}.
\end{align}

\subsection{Downlink Communications}
With the knowledge of channel parameters obtained from uplink positioning, the BS performs downlink beamforming to compensate for the dominant propagation effects before transmission. Let $\hat{{\mathbf{G}}}_{\mathrm{D}}=\mathbf{G}_\text{U}^T$ denote the downlink channel matrix reconstructed from the estimated channel parameters through the uplink transmission phase. To mitigate the channel effect, the transmit signal is multiplied by a ZF precoding matrix $\mathbf{W}$, which can be formulated as
\begin{align}\label{}
	\mathbf{W} = \gamma\hat{\mathbf{G}}_{\mathrm{D}}^H (\hat{\mathbf{G}}_{\mathrm{D}} \hat{\mathbf{G}}_{\mathrm{D}}^H)^{-1},
\end{align}
where $\gamma = \frac{\sqrt{MN}}{\|\mathbf{W}\|^2}$ is the normalization coefficient. The estimated channel parameters are used to pre-equalize the downlink channel, thereby reducing interference before transmission. Considering the LMMSE receiver, the downlink detection exploits the effective channel matrix $\mathbf{G}_\text{eff} =\mathbf{G}_{\mathrm{D}}\mathbf{W}$ with $\mathbf{G}_{\mathrm{D}}$ denoting the downlink channel matrix, which reflects both the real downlink communication channel and the precoding applied at the BS. Under this model, the received signal is expressed as
\begin{align}\label{DD channel}
	\mathbf{y} = \mathbf{G}_\text{eff} \mathbf{x} + \mathbf{n},
\end{align}
where $\mathbf{x}$ denotes the transmitted data vector and $\mathbf{n}$ represents the additive noise. Then, we exploit the LMMSE detector to recover the transmit symbols, yielding
\begin{align}\label{}
	\hat{\mathbf{x}}
	= \left( \mathbf{G}_\text{eff} ^{H}\mathbf{G}_\text{eff}  + \sigma^2\mathbf{I} \right)^{-1} \mathbf{G}_\text{eff} ^{H}
	\mathbf{y},
\end{align}
thus implying that the uplink estimated channel parameters can be leveraged in the downlink pre-equalization to enhance communication reliability under low-resolution ADCs. As a matter of fact, the accuracy of the estimated parameters influences the effectiveness of downlink beamforming.

\section{Simulation Results}\label{Section5}
In this section, we present simulation results to evaluate the overall performance of our OTFS-IPAC framework with low-resolution ADCs. The Doppler shift $k_p$ is generated according to Jakes’ model, and we assume the user moves along the horizontal axis. The channel gains are modeled as independent zero-mean complex Gaussian random variables with variances of $\xi_{p}$, which is determined by the normalized exponential power delay profile, yielding $\xi_p = \exp(-\mu l_p)/\sum_{p} \exp(-\mu l_p)$ with $\mu = 0.1$ \cite{yuan2021iterative}. Other simulation parameters are summarized in Table~\ref{parameter setting}. 

\begin{figure}[htbp]
	\centering
	\includegraphics[width=0.8\linewidth]{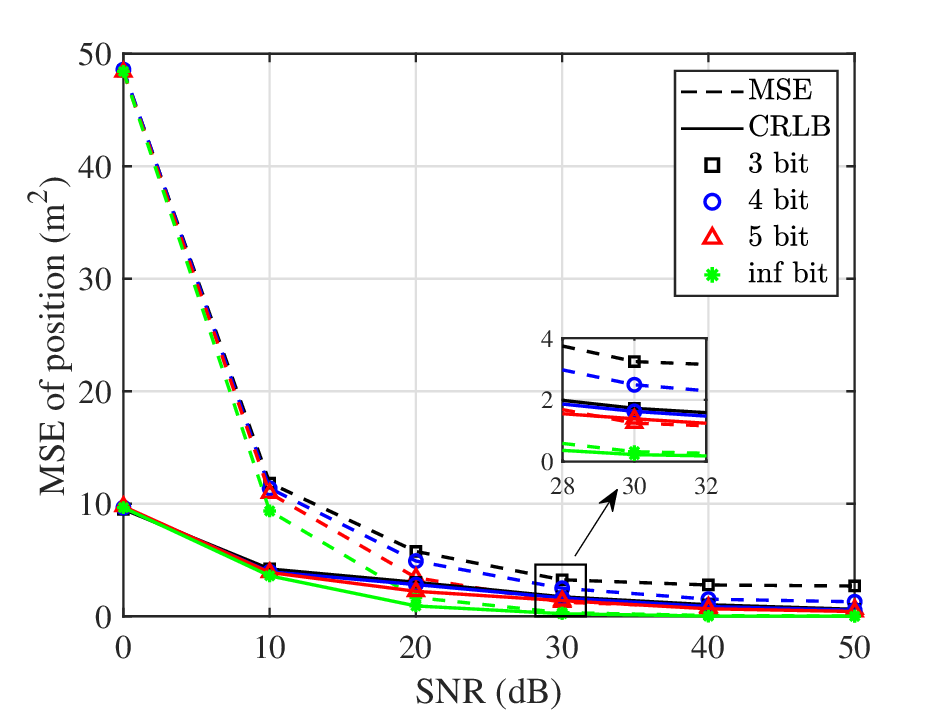}
	\caption{Uplink MSEs and CRLBs of user position versus SNR, under different quantizer resolutions of ADCs.}
	\vspace{-1em}
	\label{position}
\end{figure}
\begin{figure}[htbp]
	\centering
	\includegraphics[width=0.8\linewidth]{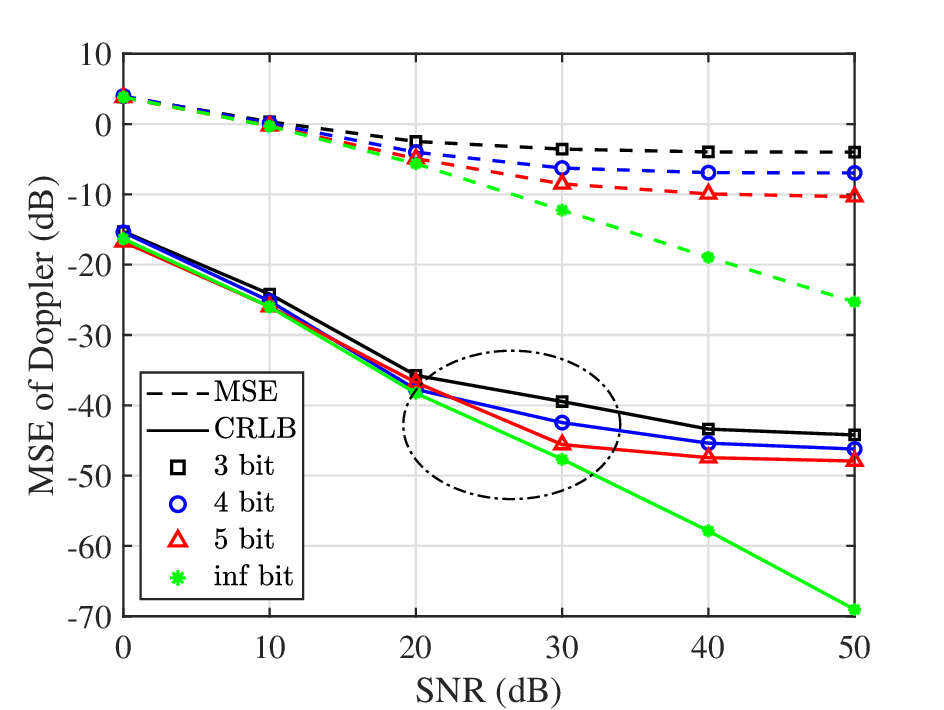}
	\caption{The MSEs and CRLBs of Doppler using different quantizer resolutions of ADCs.}
	\vspace{-1em}
	\label{doppler}
\end{figure}
\begin{figure}[htbp]
	\centering
	\includegraphics[width=0.8\linewidth]{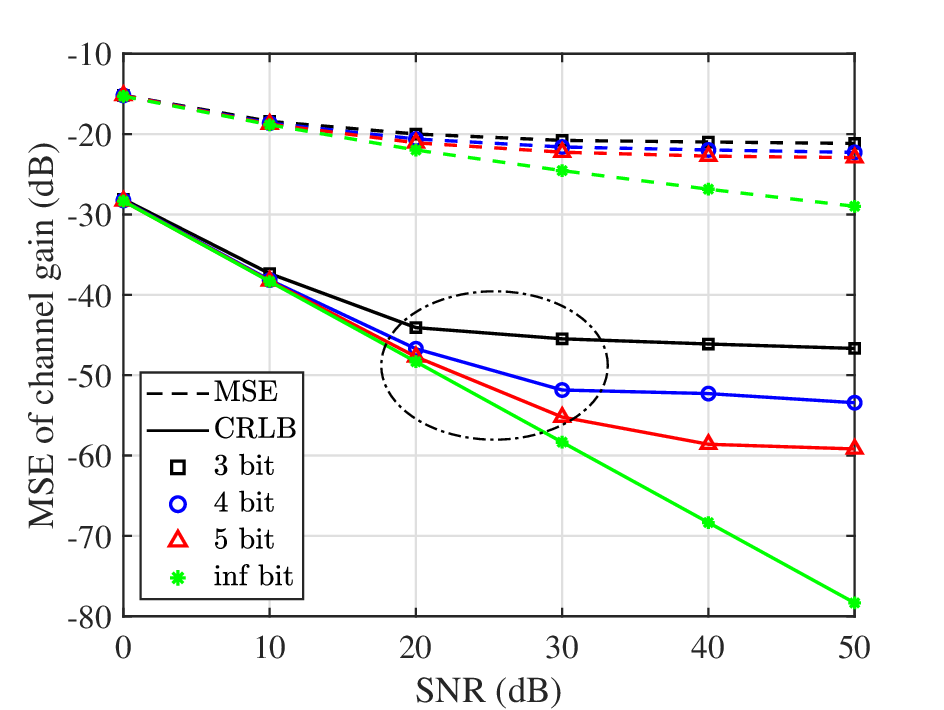}
	\caption{The MSEs and CRLBs of channel gain, where different quantizer resolutions of ADCs are exploited.}
	 \vspace{-1em}
	\label{hn}
\end{figure}

Figs. \ref{position} – \ref{hn} depict the MSE performance and the derived CRLBs as functions of the SNR for the user position, Doppler, and channel gains, respectively, under various ADC resolutions. For SNRs below 20dB, both the MSEs and CRLBs exhibit large values across all quantization resolutions, and the performance differences among various bit depths are negligible. This behavior arises because the channel noise dominates the estimation error, thereby obscuring the impact of quantization. As the SNR increases, the distortion introduced by low-resolution ADCs becomes more pronounced. When the SNR exceeds $20$ dB, the MSEs corresponding to low-resolution ADC scenarios exhibit error floors, while the associated CRLBs also remain constant beyond $30$ dB, revealing a quantization-induced performance limit. By contrast, in the infinite-resolution case, both the MSE and CRLBs decrease monotonically as the SNR increases.

\begin{table}[t!]
\footnotesize
\centering
\caption{Simulation Parameters}
\label{parameter setting}
\begin{tabular}{l c}
\hline
\textbf{Parameter} & \textbf{Value} \\
\hline
Number of subcarriers $M$ & $16$ \\
Number of OTFS symbols $N$ & $8$ \\
Number of transmit/receive antennas $N_t$, $N_r$ & 16 \\
Number of paths $P$ & 3 \\
Carrier frequency $f_c$ & $4$ GHz \\
Subcarrier spacing $\Delta f$ & $15$ kHz \\
Maximum normalized Doppler $k_{\max}$ & $0.59$ \\
Maximum velocity & $300$ km/h \\
Length of subarrays $L$ & $8$ \\
Position of the user terminal& $(883,883)$ m \\
Position of the BS & $(0,0)$ m\\
\hline
\end{tabular}
\end{table}

Specifically, we observe in Fig. \ref{position} that the gap between the estimation and CRLB becomes closer as the SNR increases. This is because the CRLB implies the achievable estimation performance regardless of the algorithm used. Moreover, with a $5$-bit ADC at an SNR of $30$ dB, the positioning MSE of the proposed method is nearly identical to the CRLB.

In Fig. \ref{doppler} and Fig. \ref{hn}, there is a performance gap between MSEs and CRLBs due to the error propagation effect of the iterative interference cancellation algorithm. Moreover, for Doppler estimation, the CRLB of $5$-bit ADCs exhibits an approximately $10$ dB SNR loss compared to the infinite-resolution ADC case, given the MSE of $-10$ dB. Similarly, in Fig. \ref{hn}, an approximately $10$ dB SNR gap is observed between the CRLBs of the $5$-bit and infinite-resolution ADC cases at an MSE level of $-58$ dB. Overall, the simulation results confirm that the performance degradation of OTFS-IPAC systems with low-resolution ADCs is most pronounced in the high SNRs.

\begin{figure}[htbp]
	\vspace{-1em}
	\centering
	\includegraphics[width=0.8\linewidth]{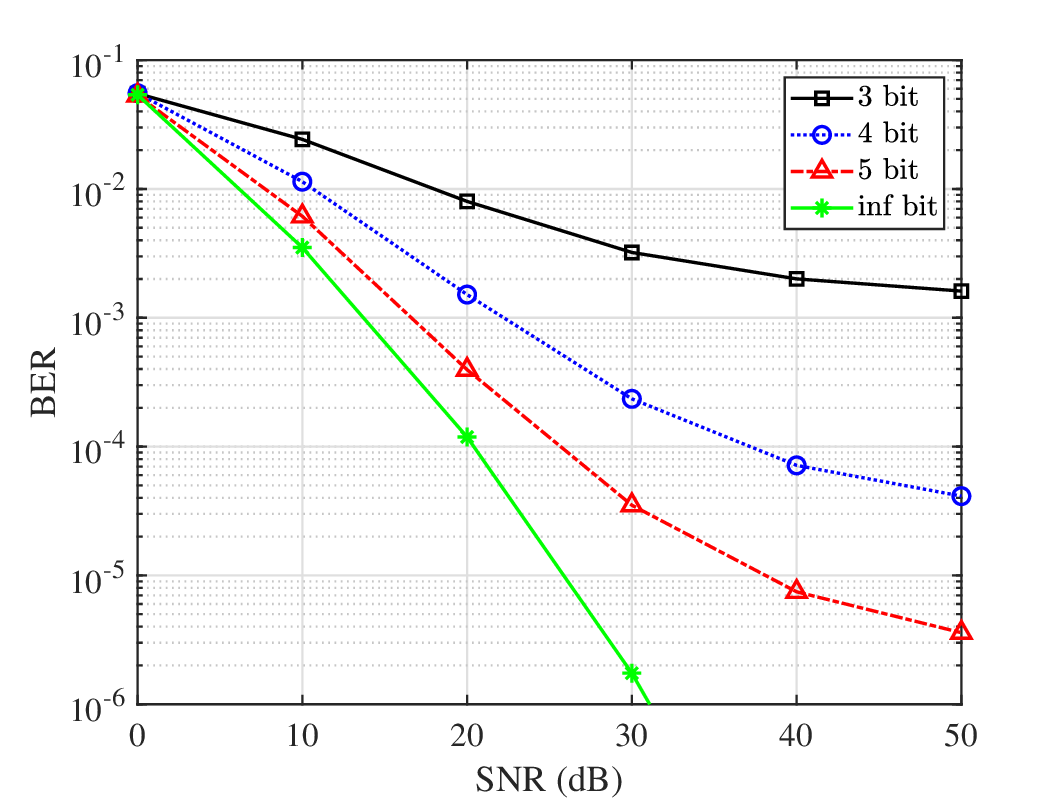}
	\caption{Downlink BER performance of the LMMSE detector with low-resolution ADCs.}
	\vspace{-1em}
	\label{BER}
\end{figure}

Fig. \ref{BER} illustrates the downlink BER performance among different ADC resolutions. We have the following observations from Fig. \ref{BER}. Firstly, given the SNRs, we can observe that increasing the quantizer resolution significantly improves the BER performance. At a target BER of $10^{-4}$, the system equipped with infinite-bit ADCs attains approximately $5$ dB and $17$ dB SNR gains compared to the $5$-bit and $4$-bit cases, respectively. Secondly, in the case of $b=3$, the BER curve becomes saturated when SNR is higher than $40$ dB. This behavior is attributed to the limited parameter estimation performance in the uplink. Consequently, the quantization distortions result in an unavoidable mismatch between the beamforming matrix and the actual channel matrix. Furthermore, we can observe that the $5$-bit scenario attains a BER of $3\times10^{-4}$ at $30$ dB. Combining the observations in Fig. \ref{position} suggests that using a $5$-bit ADC can achieve good positioning and communication performance at an SNR of $30$ dB.

\section{Conclusion}\label{Section6}
An OTFS-IPAC framework with low-resolution ADCs was proposed, in which a BS equipped with a ULA serves a single-antenna user terminal.  In our proposed framework, the channel parameter estimation and user positioning are first carried out in the uplink, while subsequent downlink beamforming is designed using the estimated uplink channel parameters. In addition, the CRLBs for channel parameters and user position were derived, serving as fundamental performance benchmarks for the proposed OTFS-IPAC system. An SS-MUSIC algorithm combined with iterative interference cancellation was further developed to estimate the relevant parameters. Simulation results have shown the impact of low-resolution ADCs on both positioning accuracy and communication performances within the proposed framework. Our study has shown that the uplink positioning and channel parameter estimation performance play a major role in affecting the downlink communication BER.

\renewcommand{\refname}{References}
\mbox{} 
\nocite{*}
\bibliographystyle{IEEEtran}
\bibliography{IPAC.bib}

\end{document}